\documentclass[journal]{IEEEtran}
\IEEEoverridecommandlockouts
\usepackage{cite}
\usepackage{amsmath,amssymb,amsfonts,extpfeil}
\usepackage{algorithmic}
\usepackage{graphicx}
\usepackage{textcomp}
\usepackage{algorithmic}
\usepackage{algorithm}
\usepackage{lettrine}

\usepackage{caption}
\makeatletter

\newcommand{\Rmnum}[1]{\expandafter\@slowromancap\romannumeral #1@}
\makeatother

\usepackage[english]{babel}
\usepackage{amsmath}
\usepackage{graphicx} 
\usepackage{epstopdf}
\usepackage{amssymb}
\usepackage{amsfonts}
\usepackage{amsthm}
\usepackage{cite}
\usepackage{bm,comment}

\usepackage{subeqnarray}
\usepackage{multicol}
\usepackage{multirow}
\usepackage{diagbox}
\usepackage{slashbox}
\usepackage{stfloats}
\usepackage{float}
\usepackage{color} 
\usepackage{cases}

\graphicspath{{figs/}}
\usepackage[caption=false,font=footnotesize,
labelfont=rm,textfont=rm]{subfig}
\renewcommand{\baselinestretch}{0.98}

\begin{document}
	
\setlength{\lineskiplimit}{0pt}
\setlength{\lineskip}{0pt}
\setlength{\abovedisplayskip}{3pt}   
\setlength{\belowdisplayskip}{3pt}
\setlength{\abovedisplayshortskip}{3pt}
\setlength{\belowdisplayshortskip}{3pt}
	
\title{Unlocking the Potential of Movable Antennas: General and Practical Antenna Position Optimization}
\author{
Weidong Mei, \IEEEmembership{Member, IEEE}, Changhao Liu, Dong Wang, Xin Wei, Yiming Wu, Boyu Ning, Zhi Chen, \IEEEmembership{Senior Member, IEEE}, Jun Fang, \IEEEmembership{Senior Member, IEEE}, Rui Zhang, \IEEEmembership{Fellow, IEEE}
	\thanks{W. Mei, C. Liu, D. Wang, X. Wei, Y. Wu, B. Ning, Z. Chen, and J. Fang are with the National Key Laboratory of Wireless Communications, University of Electronic Science and Technology of China, Chengdu 611731, China. R. Zhang is with the Department of Electrical and Computer Engineering, National University of Singapore, Singapore 117583.}
}
\maketitle
\maketitle
	
\begingroup
\allowdisplaybreaks

\begin{abstract}
Recently, movable antenna (MA) has attracted wide attention in wireless communications due to its potential in enhancing wireless communication performance via local movement within a confined region. However, antenna position optimization (APO) has emerged as a major challenge for MAs, due to the lack of a tractable, analytical, and accurate channel model in terms of antenna positions. Although existing works have developed various algorithms for APO, most of them are based on simplified theoretical channel models, which limit their generality. To address this challenge, in this article, we present more general and effective APO algorithms for different purposes, categorized as continuous APO and discrete APO, respectively. Continuous APO is mainly applied for flexible array signal processing to boost large-scale communication performance, while discrete APO is applied for small-scale multi-path channel reshaping. Specifically, the discrete APO discretizes the antenna movement region into multiple sampling points and employs discrete algorithms to determine the optimal MA positions based on the point-wise channel state information (CSI), without the need for an analytical channel model. To reduce the overhead for CSI acquisition, we also present more efficient learning-based APO algorithms that operate without requiring full point-wise CSI. Finally, we compare the application scenarios of the proposed algorithms and validate their effectiveness with numerical results.
\vspace{-6pt}
\end{abstract}
    
\begin{figure*}[t]
\centerline{\includegraphics[width=0.93\textwidth]{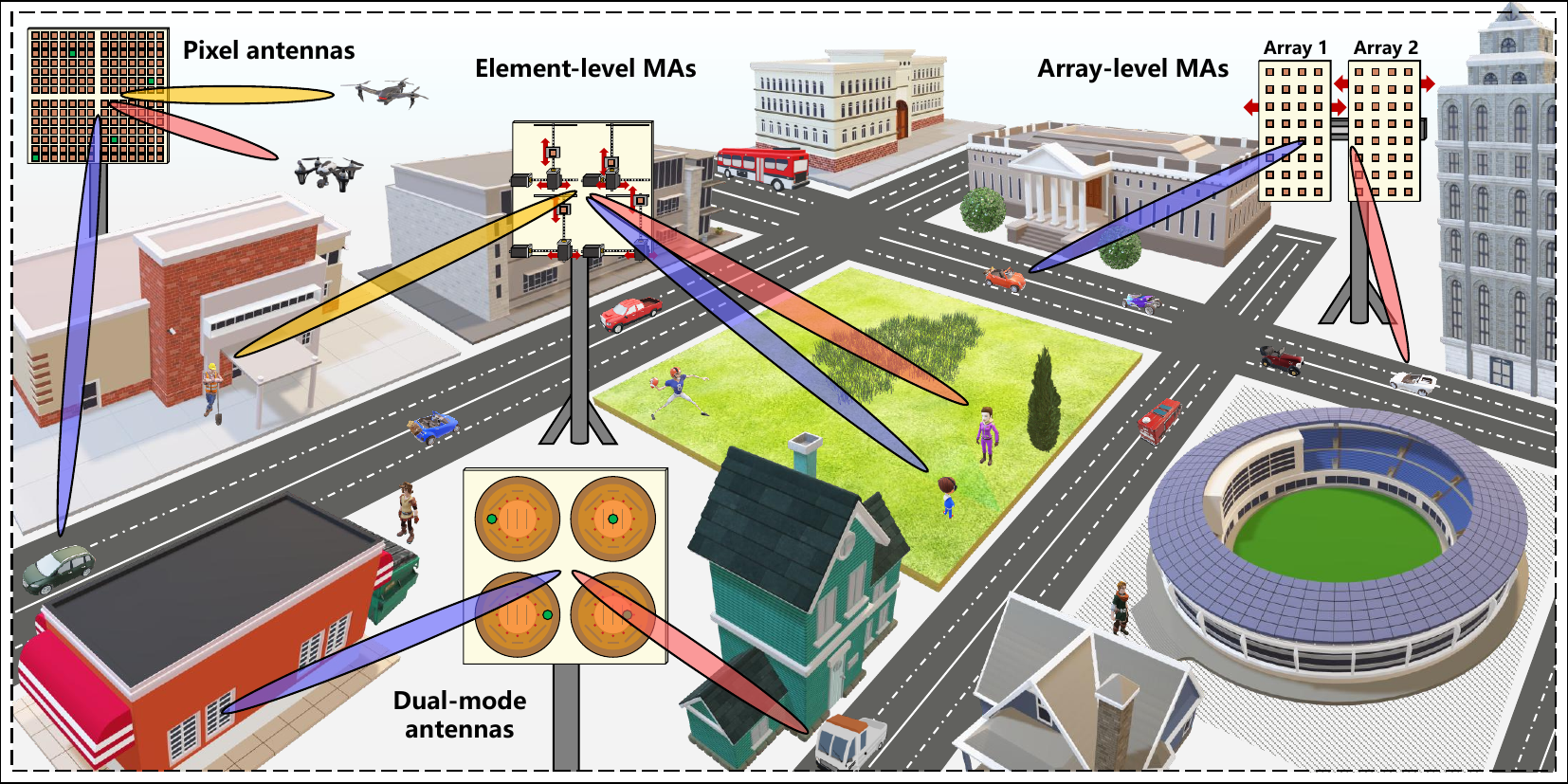}}
\caption{Illustration of practical architectures and implementations of MAs.}
\label{framework}
\vspace{-6pt}
\vspace{-6pt}\vspace{-3pt}
\end{figure*}

\section{Introduction}
The rapid advancement of next-generation applications, such as autonomous driving, low-altitude economy, and the metaverse, has driven an unprecedented demand for ultra-high data rates and ultra-low latency. To meet these requirements, the wireless research community is actively investigating high-frequency communication technologies that exploit the ultra-wide bandwidth of millimeter-wave (mmWave) and terahertz (THz) signals. In particular, extremely large antenna arrays (ELAAs) have been identified as a promising solution, which provides pronounced beamforming gains to overcome the severe path loss over mmWave/THz frequency bands.
Despite their potential, high-frequency wireless communications with ELAAs face several key challenges, such as overheating due to reduced inter-antenna spacing, limited flexibility due to fixed spatial correlation, and inability to reshape wireless channels.

Recently, movable antennas (MAs) have drawn increasing attention in the field of wireless communications. Compared to conventional fixed-position antennas (FPAs), MAs allow multiple antennas to be flexibly moved within a confined region \cite{zhu2025a}. Empowered by this additional degree of freedom (DoF), MAs can dramatically reduce the number of RF chains required in ELAAs while maintaining performance comparable to FPAs. Compared to FPAs which may experience deep fading at specific positions for a given time and/or frequency, the positions of MAs can be adjusted to circumvent such deep fading positions, effectively reshaping the wireless channel into a more favorable condition. Furthermore, by properly adjusting the antenna geometry, MAs can alter the spatial correlation among steering vectors corresponding to different angles. This allows larger inter-antenna spacing without sacrificing beam steering capability, thereby mitigating the overheating issue and enabling more flexible beamforming.  As depicted in Fig.~\ref{framework}, existing MA architectures vary in their implementation to balance control complexity and achievable performance, ranging from element-level movement to array-level movement, and from local position adjustment to global repositioning. To reduce the latency associated with mechanical antenna movement, recent studies have introduced electrical reconfiguration methods, such as pixel antennas \cite{chen2025remaa} and dual-mode antennas, which enable rapid, effective antenna repositioning in response to time-varying channel conditions.

Despite the above appealing advantages of MAs, a new antenna position optimization (APO) problem arises. From an array signal processing perspective, the transmit beamforming and antenna position become intricately coupled in shaping the spatial correlation among array response vectors, thus requiring more advanced APO algorithms to delve into their joint impacts. On the other hand, from a multi-path channel reshaping perspective, practical channels depend on complex environmental factors, making it difficult to characterize their relationship with antenna positions for APO. Although various APO algorithms have been developed in the literature, they mostly rely on theoretical and simplified continuous channel models to ease APO \cite{zhu2025a}, which may limit their generality and even cause performance degradation in practice due to the model mismatch. In addition, continuous optimization algorithms are misaligned with the practical finite-resolution constraints of antenna position adjustment (e.g., those imposed by stepper motors or discrete architectures such as pixel antennas), and they are also prone to undesired local optima due to the highly nonlinear channel structure\cite{ning2024movable}.

Motivated by the above, we present more general and effective APO algorithms in this article to accommodate diverse purposes and channel models in practice, categorized as continuous APO and discrete APO, respectively. Continuous APO algorithms optimize antenna positions as continuous variables, thus requiring an analytical channel model in terms of antenna positions, which may be difficult to attain in practice. Thus, continuous APO algorithms are more suitable for reconfiguring spatial correlation of steering vectors in the beam domain, such as beam nulling, multi-beam forming, wide-beam coverage, side-lobe suppression, etc. Discrete APO algorithms, on the other hand, are mainly applied for reconfiguring practical multi-path channels typically lacking explicit analytical forms in terms of antenna positions and relying on empirical or site-specific data. To tackle this challenge, the discrete APO discretizes the antenna movement region into multiple sampling points, thus converting the continuous APO into a discrete sampling point selection problem. As a result, discrete optimization techniques can be employed to optimize the MA positions with the point-wise CSI, dispensing with the need for an analytical channel model. To further reduce the CSI acquisition overhead, more robust APO algorithms without full CSI are also presented in this article by leveraging advanced channel learning/prediction techniques. Finally, we compare the application scenarios of the proposed algorithms and validate their effectiveness through numerical results in both beam- and channel domains. It is also worth noting that while several recent magazines have explored various aspects of MAs (e.g., \cite{sun2025movable, ning2024movable}), this work represents the first study specifically dedicated to APO for MA systems.

\begin{figure*}[t]
\centering
    \includegraphics[width=0.94\linewidth]{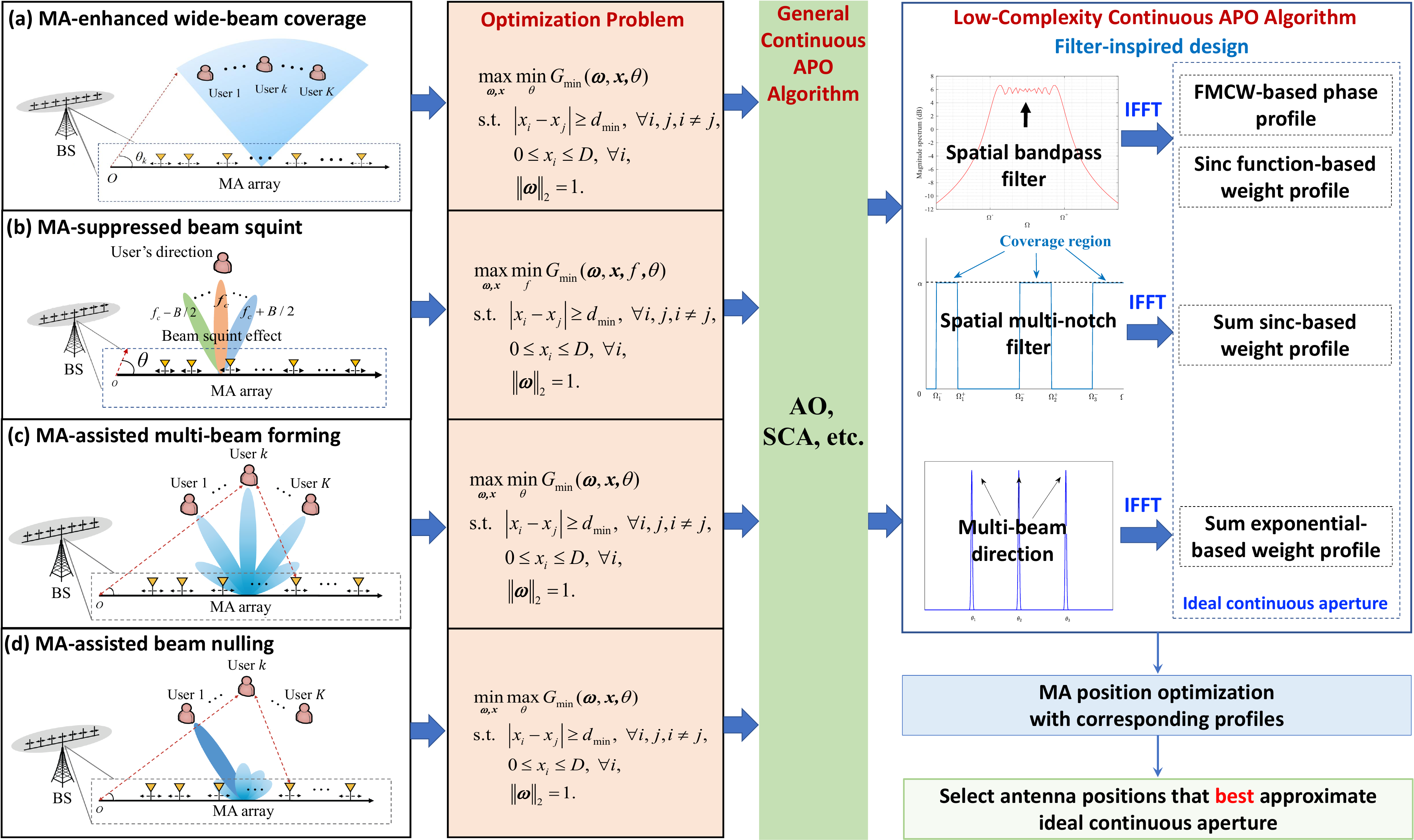}
    \caption{Illustration of the general framework of continuous APO for flexible beamforming.}
    \label{fig_beamcoverage}
\vspace{-6pt}
\vspace{-6pt}\vspace{-3pt}
\end{figure*}

\section{Continuous APO}
In this section, we present continuous APO algorithms, which are primarily based on derivatives or higher-order gradients, such as successive convex approximation (SCA), projected gradient algorithms and so on. As previously discussed, due to the difficulty in developing analytical and tractable channel models for MAs in practice, continuous APO algorithms are more suitable for reconfiguring line-of-sight (LoS)-dominant channels or so-called beam-domain channels to capture the large-scale performance enhancement. In the following, we show the applications of several continuous APO algorithms for typical beam-domain scenarios with continuous/discrete angular regions/directions.  \vspace{-6pt}

\subsection{Continuous Angular Region} 
First, we consider beam-domain optimization for continuous angular regions, such as wide-beam coverage or side-lobe suppression for given angular regions, and take 1D MA as an example. For wide-beam coverage, the associated optimization problem aims to maximize the minimum beam gain over one or multiple desired angular regions by jointly optimizing the transmit beamforming and antenna position vector (APV). The resulting optimization problem is non-convex and difficult to optimally solve due to the continuous angle range and the coupling between the transmit beamforming and the APV. Nonetheless, thanks to the analytical nature of this problem, alternating optimization (AO) can be employed to obtain a high-quality suboptimal solution by decomposing the joint optimization problem into two subproblems for beamforming and APV optimization, respectively. Each subproblem can be efficiently solved by applying the SCA technique \cite{wang2025movable}. However, such an AO algorithm generally results in high computational complexity and barely unveils the structural properties of the optimal APV. 

To tackle this issue, a filter-inspired APO framework was recently proposed by exploiting the analogy between beam gain synthesis and frequency-domain filtering (via the fast Fourier transform (FFT)). Specifically, to realize a desired beam-gain pattern, an inverse FFT (IFFT) can be performed on it to obtain an ideal continuous amplitude and/or phase profile over the antenna movement region. Based on these profiles, the APV is optimized to select the most influential antenna positions subject to inter-antenna spacing constraints. This also reveals an interesting beamforming property of APO: {\it MAs can be used to achieve the closest practical approximation to the performance of an ideal continuous aperture.} Taking wide-beam coverage over a single angular region as an example, the desired beam pattern is analogous to a band-pass filter in the angular/spatial domain. Accordingly, the corresponding amplitude and phase profile can be obtained via IFFT as a sinc-shaped function (modulated by a cosine term). Moreover, if the beamforming is restricted to analog beamforming with a constant amplitude, the phase profile can be obtained as the frequency modulation continuous wave (FMCW) chirp signals\cite{wang2025movable}. Given these profiles, only antenna position remains to be optimized, which thus incurs a much lower complexity order compared to the joint beamforming and position optimization. In addition, for wide-beam coverage over multiple angular regions, the desired beam-gain pattern resembles a multi-notch filter, for which the associated amplitude and phase profile can be obtained via IFFT as a sum of sinc functions\cite{wang2026movable}.

It is worth further noting that the wide beam coverage can also be achieved beyond the spatial domain. For example, in terahertz (THz) communication systems, due to the much larger signal bandwidth, ``beam squint'' effect may occur, resulting in more significantly varying beam gain over the entire frequency band. In this case, wide beams are also needed in the frequency domain to mitigate the beam squint effects by generating uniform beam gains over the entire frequency band, for which MAs can be employed. In addition to wide-beam coverage, continuous APO can also be used to effectively suppress the beam gain within a single or multiple angular regions, e.g., side-lobe level suppression. The associated problem can be formulated as minimizing the maximum beam gain over these regions while ensuring the beam gain within another region, which deserves further investigation in future work. Similarly, this new problem can be effectively solved via AO and SCA-based algorithms as well.\vspace{-6pt}
  
\subsection{Discrete Angles} 
Beam-domain optimization with discrete angles usually pertains to multi-beam forming and beam nulling. Multi-beam forming aims to achieve a high beam gain over several desired discrete angles. The corresponding  optimization problem aims to maximize the minimum beam gain over all desired directions by jointly optimizing beamforming and APV. Although this optimization problem shares similarity to that for wide-beam coverage, their design principles are different. For example, in the case of two antennas, it can be shown that a larger/smaller antenna aperture is desired to achieve multi-beam forming and wide-beam coverage, respectively. Their difference can also be interpreted from the filter design perspective. For multi-beam forming, the desired beam-gain pattern can be obtained by reducing the pass bands for continuous angular regions to impulses for discrete angles. As such, we can reconstruct the amplitude and phase profile as the sum of complex exponential functions, based on which the APV is optimized. For the max-min problem, it can also be suboptimally solved by invoking the gradient-based algorithms such as SCA \cite{wang2025movable}. It was also shown in \cite{zhu2025a} that in the case of an arbitrarily large antenna movement region, the optimal APV can be obtained in closed-form. 
  
On the other hand, beam nulling aims to jointly optimize the APV and the beamforming vector to maximize the beam gain at a desired target while nulling interference toward undesired targets. Similar to the multi-beam forming scenario, the optimal APV can also be obtained in closed-form under the assumption of an arbitrarily large antenna movement region, which is constructed by using prime factorization of the number of nulls \cite{zhu2025a}. However, developing efficient continuous APO algorithms in other general scenarios remains challenging for beam nulling. In particular, under zero-forcing beamforming, the objective function admits a complicated (often highly nonlinear) expression, which makes gradient-based optimization computationally inefficient and poorly conditioned despite its analytical nature. In this case, discrete APO algorithms may be invoked to circumvent this difficulty, as addressed next. The presented continuous APO algorithms in this section for MAs in the beam domain are illustrated in Fig.~\ref{fig_beamcoverage}.\vspace{-6pt}
  
\begin{figure*}[t!]
\centerline{\includegraphics[width=0.93\textwidth]{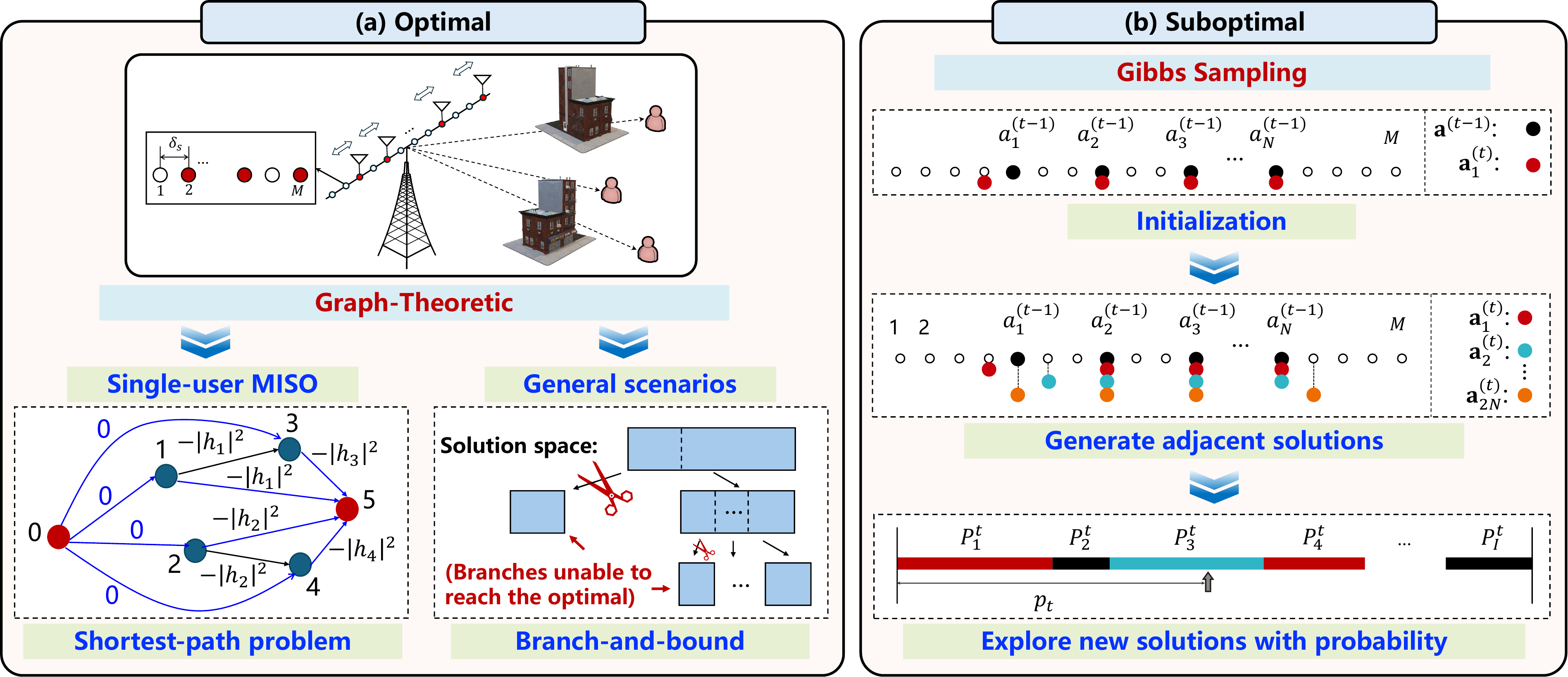}}
\vspace{-2pt}
\caption{Illustration of the general framework of discrete APO algorithms.}
\label{topic_2}
\vspace{-6pt}
\vspace{-6pt}
\vspace{-2pt}
\end{figure*}

\section{Discrete APO}
As previously mentioned, continuous APO may face a variety of issues such as undesired local optima and model mismatch in the presence of complex expressions (e.g., practical multi-path channels or beam nulling). To address these issues, in this section, we present various discrete algorithms to optimize the MA positions. Different from the continuous APO, the discrete APO discretizes the antenna movable region into multiple sampling points and aims to select an optimal set of antenna positions from these sampling points with their point-wise CSI. Such CSI can be obtained either via interpolation or extrapolation from channel estimation at partial positions \cite{zhang2025successive} or sampling based on an estimated/known continuous channel map \cite{xiao2024channel}. \vspace{-6pt}

\subsection{Optimal Algorithms}
First, for a simple point-to-point multi-input single-output (MISO) system, a graph-theoretic algorithm can be utilized to find globally optimal MA positions in polynomial time\cite{mei2024movable}. Specifically, in the single-user setup, the transmit beamforming can be obtained as the maximum transmission ratio (MRT). Hence, the received signal power at the user becomes proportional to the sum of channel power gains at the selected positions, which allows for an efficient graph-based problem reformulation. As shown in Fig.~\ref{topic_2}, a graph is constructed, with its vertex set being the set of all sampling points. Moreover, an edge exists between two vertices if and only if their corresponding two sampling points satisfy the constraints associated with MAs (e.g., inter-MA spacing constraints for avoiding mutual coupling). By further introducing two dummy vertices and properly setting the edge weights, the original point selection problem becomes equivalent to a {\it fixed-hop} shortest-path problem\cite{mei2024movable}, which can be optimally solved via dynamic programming (DP) with a quadratic complexity order of $\mathcal{O}(M^2N)$, with $M$ and $N$ denoting the number of sampling points and the number of transmit antennas, respectively.

While the above graph-theoretic approach provides an efficient and globally optimal solution for the single-user case, it cannot be directly extended to other more general scenarios. This is due to the more complex coupling between antenna positions and other design variables, as well as more complex design objectives and constraints, making the discrete point selection problem NP-hard. To address this issue, a higher-complexity \textit{branch-and-bound} (BnB) method can be employed to obtain global optima via solution enumeration\cite{mei2024secure}.

As depicted in Fig.~\ref{topic_2}, the crux of the BnB method is to divide the feasible solution space into partitions arranged in a tree structure. Each node of the BnB tree represents  a subset of the feasible set. By subdividing the feasible region into a series of subsets, the problem is decomposed into a series of manageable subproblems corresponding to the subsets. For each subproblem, a lower- and an upper-bound solution is constructed and used for solution pruning. For example, by jointly implementing BnB and the graph-theoretic algorithm, the globally optimal MA positions have been found in the scenarios of physical-layer security \cite{mei2024secure}, where the shortest-path algorithm can be utilized to characterize a performance upper bound on the incumbent solution for pruning.
Although the worst-case complexity of the BnB method is exponential in the number of antennas, the use of effective bounding and pruning strategies often significantly reduces the number of searches required in practice, making it feasible for moderate-sized problems.\vspace{-6pt}

\subsection{Suboptimal Algorithms}
To strike a practical balance between performance and complexity, several efficient suboptimal discrete APO algorithms have been proposed. For example, a straightforward method is to perform a sequential update of the optimal sampling point for each MA (with the positions of all other MAs fixed) over multiple rounds. The merits of this sequential update are threefold. First, it is provably convergent, as it yields monotonically increasing/decreasing objective values in updating the positions of the MAs. Second, the sequential update operations can be performed regardless of the specific problem structure and also dispense with complex gradient calculations as required in continuous APO. Third, it incurs a low computational complexity in linear order, i.e., $\mathcal{O}(MN)$.
However, despite the above merits, the sequential update may suffer from an undesired local optimum in the iteration.

To avoid convergence to low-quality local optima, some solution exploration schemes, such as Gibbs sampling, can be utilized between two consecutive rounds of the sequential update. The core idea is to explore more feasible solutions within the whole solution space. To this end, a set of random and adjacent feasible solutions are generated, given the current converged solution by sequential update. All candidate APV solutions (both adjacent and random) are assigned different selection probabilities to balance exploration during the search, as depicted in Fig.~\ref{topic_2}. Random candidates help prevent the iterations from getting trapped in locally optimal configurations and reduce sensitivity to initialization. Adjacent candidates, in contrast, enable efficient, low-cost refinement by exploiting local structure. By combining these two types of candidate solutions, the algorithm achieves a better exploration–exploitation tradeoff, which in turn improves overall solution quality. It should also be noted that the Gibbs sampling incurs a linear complexity order. Hence, the complexity order of the overall sequential update process with GS sampling remains linear \cite{liu2026a}. The above presented discrete APO algorithms for MAs are illustrated in Fig.~\ref{topic_2}.\vspace{-6pt}

\begin{figure*}[t!]
    \centering
    \includegraphics[width=0.85\linewidth]{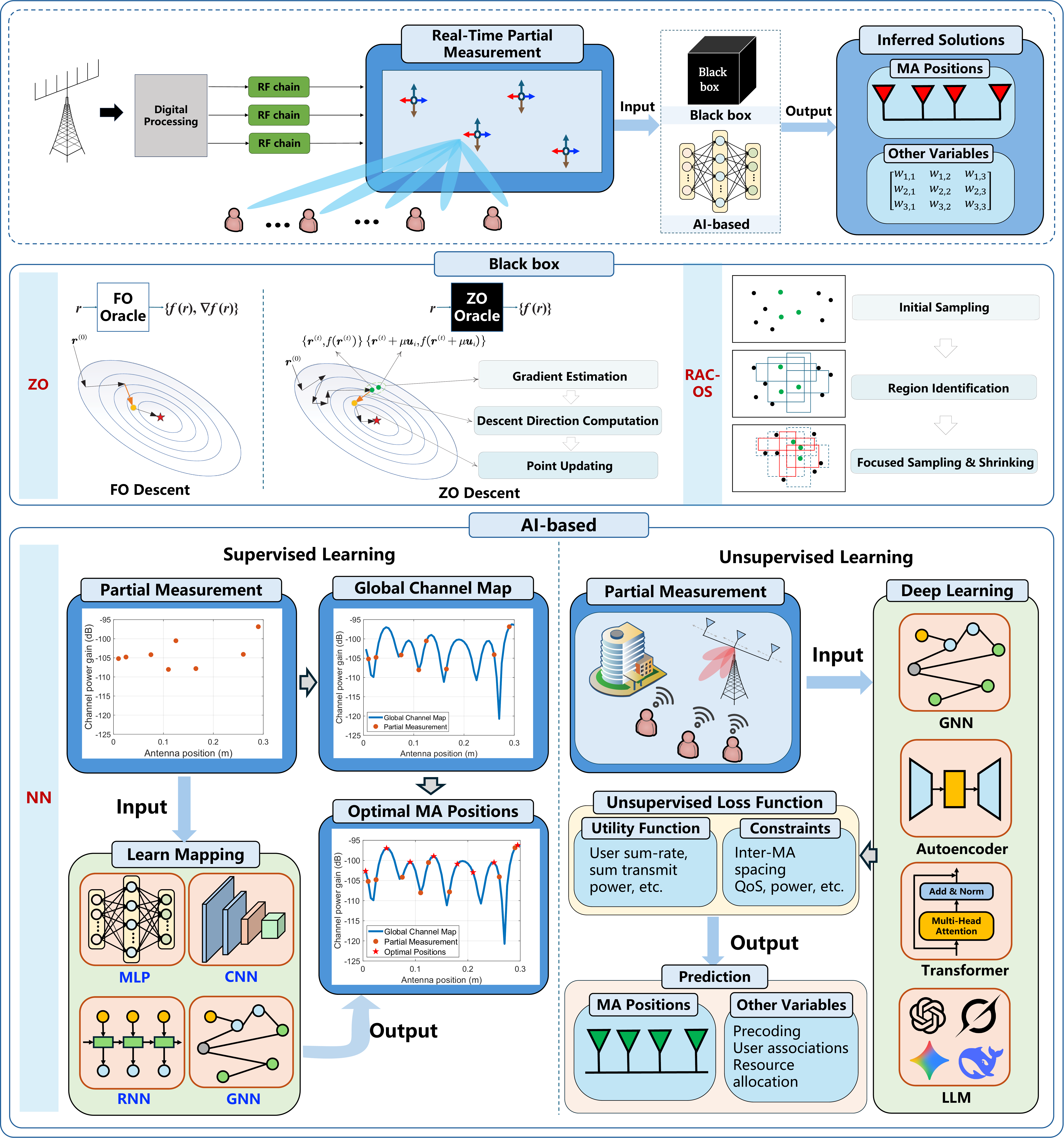}
    \caption{Illustration of the general framework of APO algorithms with partial CSI.}
   \label{topic_3}
\vspace{-6pt}
\vspace{-6pt}\vspace{-6pt}
\end{figure*}

\section{Practical APO with Partial CSI}    
Despite the effectiveness of the discrete APO, a key challenge lies in the acquisition of point-wise CSI. Unlike conventional FPA systems that only require CSI associated with the static locations of the antennas, optimizing MA positions requires CSI associated with all sampling points in the entire movement region, resembling a {\it channel map}. Acquiring a full channel map via exhaustive measurement is often impractical, while other model- and data-driven channel map estimation methods may incur high overhead or insufficient accuracy. This thus motivates the use of more practically efficient and robust APO algorithms that operate with only partial CSI, as presented in this section.\vspace{-8pt}

\subsection{Black-Box Optimization}
When explicit channel models and gradients are unavailable, black-box optimization provides a powerful alternative. In particular, the wireless system is treated as an oracle: we input an antenna configuration and observe the resulting link performance. By iteratively querying this oracle, we can navigate towards improved solutions without the need for complete/explicit CSI.

A prominent strategy in this domain is zeroth-order (ZO) methods \cite{zeng2025csi}. Inspired by gradient descent, ZO methods approximate the search direction without requiring analytical gradients. The core mechanism involves probing the system: by slightly perturbing the current antenna position and measuring the change in the output performance metric (e.g., received power), a stochastic estimate of the gradient is constructed. Advanced variants like ZO-adaptive momentum method (AdaMM) incorporate momentum and adaptive step sizes to accelerate convergence and stabilize the search in noisy environments. The primary strength of ZO optimization lies in its sampling efficiency. It can often find high-quality solutions with a relatively low number of pilot transmissions by efficiently exploring the local landscape.

However, ZO methods face inherent limitations. Their gradient-approximation mechanism assumes a relatively smooth objective landscape in a continuous domain, making them less effective for discrete position selection. More critically, they struggle with complex system constraints, such as strict interference limits or minimum distance requirements between antennas. Incorporating these constraints typically requires sophisticated penalty formulations, and the noisy gradient estimates can lead to frequent constraint violations and slow convergence.

To address these issues, especially for problems with discrete choices or intricate constraints, classification-based optimization frameworks like the randomized coordinate shrinking (RACOS) algorithm offer a complementary approach\cite{yu2016derivative}. RACOS operates on a different principle: it treats the optimization as a process of iteratively learning the promising regions of the solution space. It starts by randomly sampling a population of antenna configurations (e.g., positions and other variables like transmit power) and evaluating their performance. It then learns a simple, axis-aligned classifier to distinguish ``good'' solutions (those yielding high performance and satisfying constraints) from ``bad'' ones. In subsequent iterations, new candidate solutions are preferentially sampled from the region identified as promising by the classifier, while the search space is gradually refined. This method excels in handling combinatorial and constrained problems since constraints can be directly embedded into the definition of a ``good'' solution during the classification step. As a result, the RACOS algorithm is well-suited to discrete APO with discrete candidate antenna positions, thus deserving more investigations in future work.\vspace{-8pt}

\subsection{Artificial Intelligence (AI)-based Optimization}
While the black-box optimization eliminates the need for full channel maps, the reliance on extensive real-time measurements often introduces a high latency. To enable real-time APO with fewer measurement overhead, AI-based optimization has emerged as a promising solution. The core rationale lies in the spatial correlation of wireless channels. Specifically, in a given environment, the channel responses at different positions are not independent but are governed by the same scatterers. This implies that channel measurements at a sparse subset of sampling points contain latent information regarding the global channel map. Since the optimal antenna position is deterministically determined by this global map, it follows that {\it a direct mapping exists linking the partial measurements to the optimal MA positions.} However, deriving an explicit mathematical expression for this end-to-end mapping is intractable due to the highly non-linear coupling between antenna positions. Consequently, deep learning (DL) emerges as a natural solution to approximate this implicit mapping. The essence of this framework is to shift the heavy measurement computational cost to an offline training stage, thereby reducing the real-time operation to rapid inference only. Depending on the availability of optimal solutions during training, this framework can be divided into the following two paradigms:
	\begin{itemize}
		\item \textbf{Label-Driven Supervised Learning:} In some simple scenarios where optimal MA positions are available (e.g., single-user MISO systems), supervised learning is viable. In the offline stage, the DNN acts as \textit{a function approximator that learns the optimal APO algorithm} to map partial implicit CSI (e.g., received power) directly to the optimal positions, utilizing offline-calculated optimal MA positions (e.g., via the graph-theoretic algorithm) as ground-truth labels. Once trained, the model is deployed for real-time inference, which directly outputs the optimal MA positions based on real-time measurements \cite{lu2025learning}.
		
		\item \textbf{Objective-Driven Unsupervised Learning:} In complex scenarios like multi-user systems, obtaining ground-truth labels such as MA positions and pre-coding matrices is computationally prohibitive. To address this, unsupervised learning can be adopted, where network weights are optimized offline by directly maximizing the system's utility function (e.g., sum-rate). By incorporating the utility function and constraints (e.g., minimum inter-antenna spacing) directly into the loss function, the model learns to coordinate interference without explicit supervision. Similar to the supervised case, the real-time stage requires only a forward propagation of the pre-trained model to infer near-optimal MA positions.
	\end{itemize}

\begin{figure*}[t!]
\centering
\includegraphics[width=0.93\linewidth]{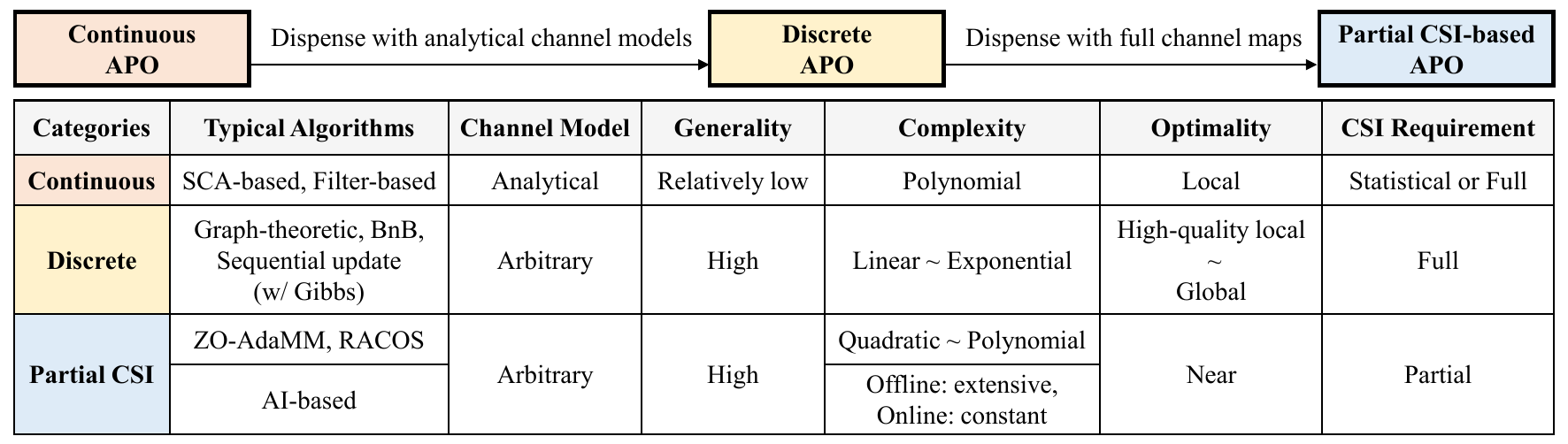}
\vspace{-6pt}
\caption{Comparison among presented APO algorithms.}
\label{table_1}
\vspace{-6pt}
\vspace{-6pt}
\vspace{-3pt}
\end{figure*}

While the above frameworks have proven effective, the rapid evolution of AI offers new paradigms to further facilitate the APO. For example, the above methods primarily focus on static optimization. However, in mobile scenarios where users or scatterers move rapidly, static solutions become obsolete quickly. By leveraging deep reinforcement learning (DRL), we can create DRL agents that formulate the dynamic APO as a sequential decision-making process. By continuously collecting information from the environment (e.g., user's location and velocity) and observing reward feedback (e.g., temporal variations in SNR), DRL can enable autonomous, real-time antenna movement tracking without repetitive re-optimization from scratch. Moreover, the accuracy of APO algorithms discussed previously heavily relies on the quality of the estimated global channel map. Generative AI, particularly diffusion models, has demonstrated exceptional capability in data generation and completion. These models can be employed to reconstruct high-precision global channel maps from extremely sparse pilot measurements, effectively resolving the problem of high movement cost that limits existing MA channel estimation techniques. Inspired by the above discussions, AI-driven APO with partial CSI is a promising direction worthy of further research. The above presented APO approaches with partial CSI are illustrated in Fig.~\ref{topic_3}.\vspace{-6pt}

\section{Comparison and Guidance}
The aforementioned APO algorithms possess distinct advantages and weaknesses in various aspects, as discussed below. \vspace{-6pt}\vspace{-6pt}\vspace{-6pt}\vspace{-3pt}
\subsection{Comparison and Discussion}
\subsubsection{Channel Model Requirement and Generality}
Continuous APO algorithms require an analytical channel model, which limits their generality. For discrete and partial CSI-based APO algorithms, they can be applied to more practical channel models with point-wise CSI or real-time inference, thus achieving a broad generality. 
\subsubsection{Computational Complexity and Optimality}
Continuous APO algorithms generally incur a polynomial complexity order to solve an analytical but nonlinear optimization problem, and a locally optimal solution can be ensured by employing various gradient-based algorithms. The discrete APO algorithms offer a good performance-complexity trade-off, achieving high-quality local or global optimality from linear to exponential complexity orders. For APO algorithms with partial CSI, near-optimal performance may be achieved given sufficient channel measurements or offline training. Among them, AI-based solutions shift most computational burden from real-time operation to an offline training phase, yielding a nearly constant, low real-time complexity.
\subsubsection{CSI Requirement}
From a beam domain perspective, continuous APO algorithms primarily target large-scale performance enhancement and therefore rely solely on statistical CSI (e.g., angle information for beam manipulation). However, when real-time performance is accounted for, continuous APO algorithms also demand full and instantaneous parameters for analytical channel models (e.g., field-response channel model\cite{zhu2025a}). Furthermore, discrete APO algorithms require full point-wise CSI or channel map, while partial-CSI-based APO algorithms avoid full point-wise CSI by leveraging prediction and inference mechanisms. \vspace{-6pt}

\subsection{Guidance and Recommendation}
To select an appropriate APO algorithm, we offer the following guidance and recommendations according to different scenarios. 
\begin{itemize}
\item For scenarios demanding theoretical analysis, e.g., in the beam domain, continuous APO algorithms are the most suitable choice by allowing for analytical channel expressions and clear performance bounds.
\item When the priority is practical implementation in extremely complex real-world environments, discrete APO algorithms are more suitable. These algorithms inherently facilitate hardware deployment, offering moderate computational complexity and robustness to channel models.
\item In scenarios constrained by limited pilot overhead, APO algorithms with partial CSI present an efficient alternative. They circumvent the need for complete CSI acquisition, which is particularly valuable for real-time optimization when measurement latency is critical, at the cost of online measurements for black-box optimization and extensive offline training for AI-based solutions.\vspace{-6pt}
\end{itemize}

\begin{figure*}[t]
\centering
\subfloat[Beam domain: wide-beam coverage]{
\includegraphics[width=0.34\textwidth]{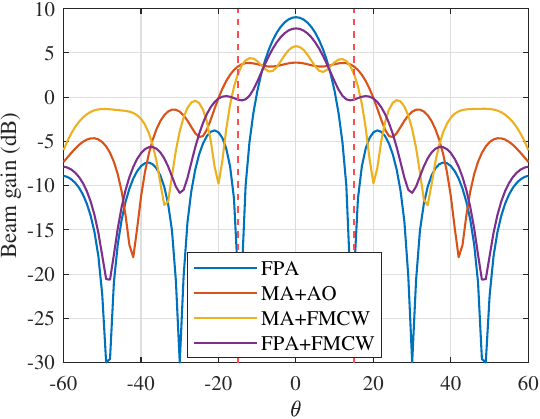}
\label{fmcw_fig}
}
\hfil
\subfloat[Channel domain: single-user MISO]{
\includegraphics[width=0.36\textwidth]{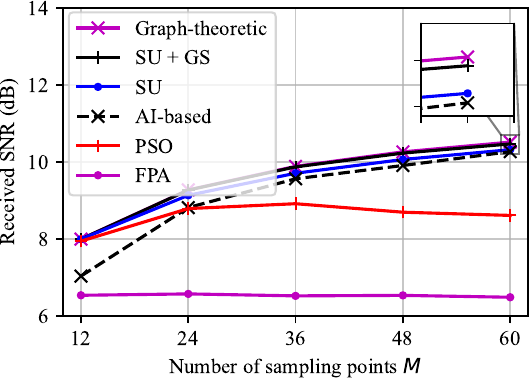}
\label{sin_use}
}
\vspace{-4pt}
\caption{Performance evaluation of presented algorithms in both beam and channel domains.}
\label{with_CSI}
\vspace{-6pt}
\vspace{-6pt}
\vspace{-3pt}
\end{figure*}

\section{Performance Evaluation}
To evaluate the performance of our presented APO algorithms, we provide numerical results to evaluate their performance in both beam and channel domains, as shown in Figs.~\ref{with_CSI}(a) and \ref{with_CSI}(b), respectively. First, we show the beam patterns achieved by different algorithms for wide-beam coverage in Fig.~\ref{with_CSI}(a). In particular, we consider a linear MA array with 8 antennas. The carrier frequency and the desired angular region are set to 6.775 GHz and $\mathcal{R}=[-15^{\circ},15^{\circ}]$, respectively. It is observed that for FPAs without a beamforming design, deep nulls occur within the target angular region $\cal R$, thus dramatically impairing the coverage performance. If the beamforming weights can be optimized for FPAs, the max-min beam gain within $\cal R$ is observed to increase to approximately 0 dB for FPAs. In contrast, thanks to APO, the max-min beam gains for MAs are observed to increase by 4 dB compared with FPAs. Particularly, it is observed that ``MA+FMCW'' achieves a close performance to ``MA+AO'', which validates the efficacy of the filter-inspired APO.

Next, we consider a single-user MISO communication system, with the carrier wavelength set to $\lambda=0.06$ meter (m). The side length of the movement region and the number of MAs are set to $A = 6\lambda$ and $N=8$, respectively. Fig.~\ref{with_CSI}(b) plots the received SNR versus the number of sampling points ($M$) within the antenna movement region under different schemes with full/partial CSI. In addition to the algorithms presented in this article, the universal albeit heuristic particle swarm optimization (PSO) algorithm is also presented as a benchmark. Note that the graph-theoretic algorithm can achieve globally optimal performance in the considered single-user setup. Hence, it is observed to outperform all other algorithms over the whole range of $M$ considered and can be employed as a reference to evaluate the optimality gaps of other algorithms. It is also observed that the sequential update algorithm with GS outperforms its counterpart without GS as well as the PSO algorithm. Moreover, its performance gap with the optimal graph-theoretic approach is negligible. These observations demonstrate the near-optimality of the sequential update algorithm with GS and the necessity of the GS procedure. Furthermore, it is observed that the AI-based approach can achieve a close performance to the optimal graph-theoretic algorithm as well. Particularly, only $50\%$ point-wise CSI is used to achieve this performance, instead of full CSI required in other algorithms.\vspace{-9pt}

\section{Conclusion}\vspace{-3pt}
In this article, we provided an overview of general and practical APO algorithms for MA systems. Continuous APO algorithms primarily target beam-domain optimization in an analytical form, which is well suited to gradient-based methods and filter-inspired design principles. In contrast, discrete APO algorithms can be applied to arbitrary channel models and system setups. Suboptimal algorithms such as sequential update with Gibbs sampling can strike a good balance between performance and complexity, avoiding the high complexity of BnB. Moreover, useful tools such as ZO, RACOS, and AI further dispense with the need for full CSI to obtain near-optimal antenna positions. The effectiveness of our presented algorithms was compared and demonstrated by numerical results. It is worth mentioning that most of the APO algorithms presented in this article can also be applied to other advanced antenna architectures\cite{shao2026a}, including six-dimensional (6D) MAs, rotatable antennas, pinching antenna systems (PASS), and fluid antenna systems (FAS). It is hoped that this article will serve as a useful reference for future research and development on these emerging reconfigurable antenna technologies.\vspace{-6pt}

\renewcommand{\baselinestretch}{0.98}
\bibliographystyle{IEEEtran}
\bibliography{mybib}

@ARTICLE{shao2026a,
  author={Shao, Xiaodan and others},
  journal={IEEE Commun. Surveys Tuts.}, 
  title={A Tutorial on Six-Dimensional Movable Antenna for {6G} Networks: Synergizing Positionable and Rotatable Antennas}, 
  year={2026},
  volume={28},
  pages={3666-3709}
}

@conference{lu2025learning,
    author = {Lu, Lele and Mei, Weidong and Wei, Xin and Hua, Haocheng and Chen, Zhi and Ning, Boyu},
    title = {Learning-Based Movable-Antenna Position Optimization with Implicit {CSI}}, 
    booktitle={Proc. IEEE Int. Symp. Person. Indoor Mobile Radio Commun.}, 
    year = {2025},
    pages={1-6}
}

@conference{yu2016derivative,
  author = {Yu, Yang and others},
  booktitle = {Proc. AAAI Conf. Artif. Intell.},
  title = {Derivative-Free Optimization via Classification},
  year = {2016},
  volume = {30},
  number = {1},
  pages = {2286-2292}
}

@ARTICLE{zeng2025csi,
  author={Zeng, Xianlong and others},
  journal={IEEE Wireless Commun. Lett.}, 
  title={{CSI}-Free Position Optimization for Movable Antenna Communication Systems: A Derivative-Free Optimization Approach}, 
  year={2025},
  volume={14},
  number={1},
  pages={53-57},
  month=jan}

@ARTICLE{liu2026a,
  author={Liu, Changhao and Mei, Weidong and Chen, Zhi and Fang, Jun and Ning, Boyu},
  journal={IEEE Wireless Commun. Lett.}, 
  title={A General Optimization Framework for Movable Antenna Systems via Discrete Sampling}, 
  year={2026},
  volume={15},
  pages={475-479}
}

@conference{mei2024secure,
  author={Mei, Weidong and Wei, Xin and Liu, Yijie and Ning, Boyu and Chen, Zhi},
  booktitle={Proc. IEEE Global Commun. Conf.}, 
  title={Movable-Antenna Position Optimization for Physical-Layer Security via Discrete Sampling}, 
  year={2024},
  location={Cape Town, South Africa},
  pages={4914-4919}
}

@article{mei2024movable,
  author   = {Mei, Weidong and Wei, Xin and Ning, Boyu and Chen, Zhi and Zhang, Rui},
  journal  = {IEEE Wireless Commun. Lett.},
  title    = {Movable-Antenna Position Optimization: A Graph-Based Approach},
  year     = {2024},
  volume   = {13},
  number   = {7},
  pages    = {1853-1857}
}

@ARTICLE{xiao2024channel,
  author={Xiao, Zhenyu and Cao, Songqi and Zhu, Lipeng and Liu, Yanming and Ning, Boyu and Xia, Xiang-Gen and Zhang, Rui},
  journal={IEEE Trans. Wireless Commun.}, 
  title={Channel Estimation for Movable Antenna Communication Systems: A Framework Based on Compressed Sensing}, 
  year={2024},
  volume={23},
  number={9},
  pages={11814-830}
}

@ARTICLE{zhang2025successive,
  author={Zhang, Zijian and Zhu, Jieao and Dai, Linglong and Heath, Robert W.},
  journal={IEEE Trans. Wireless Commun.}, 
  title={Successive {Bayesian} Reconstructor for Channel Estimation in Fluid Antenna Systems}, 
  year={2025},
  volume={24},
  number={3},
  pages={1992-2006}
}

@ARTICLE{wang2026movable,
  author={Wang, Dong and Mei, Weidong and Chen, Zhi and Ning, Boyu},
  journal={IEEE Wirel. Commun. Lett.}, 
  title={Movable Antenna Enhanced Multi-Region Beam Coverage: A Multi-Notch-Filter-Inspired Design}, 
  year={2026},
  volume={15},
  number={},
  pages={1320-1324}
}

@article{wang2025movable,
  author={Wang, Dong and Mei, Weidong and Ning, Boyu and Chen, Zhi and Zhang, Rui},
  journal={IEEE Trans. Wirel. Commun.}, 
  title={Movable Antenna Enhanced Wide-Beam Coverage: Joint Antenna Position and Beamforming Optimization},
  year={2026},
  volume={25},
  pages={3541-58},
}

@ARTICLE{sun2025movable,
  author={Sun, Chen and Li, Guyue and Wang, Shuai and Wen, Guanghui},
  journal={IEEE Wirel. Commun.}, 
  title={Movable Antenna Enabled Dual-Scale Beams for Resilient Communication, Sensing, and Security}, 
  year={2026},
  volume={33},
  number={1},
  pages={135-142}
}

@ARTICLE{ning2024movable,
  author={Ning, Boyu and Yang, Songjie and Wu, Yafei and Wang, Peilan and Mei, Weidong and Yuen, Chau and Bjornson, Emil},
  journal={IEEE Wireless Commun.}, 
  title={Movable Antenna-Enhanced Wireless Communications: General Architectures and Implementation Methods}, 
  year={2025},
  volume={32},
  number={5},
  pages={108-116}
}

@ARTICLE{chen2025remaa,
  author={Chen, Kangjian and Qi, Chenhao and others},
  journal={IEEE Trans. Commun.}, 
  title={{REMAA}: Reconfigurable Pixel Antenna-Based Electronic Movable-Antenna Arrays for Multiuser Communications}, 
  year={2025},
  volume={73},
  number={11},
  pages={12913-928}
}

@ARTICLE{zhu2025a,
  author={Zhu, Lipeng and Ma, Wenyan and Mei, Weidong and Zeng, Yong and Wu, Qingqing and Ning, Boyu and Xiao, Zhenyu and Shao, Xiaodan and Zhang, Jun and Zhang, Rui},
  journal={IEEE Commun. Surveys Tuts.}, 
  title={A Tutorial on Movable Antennas for Wireless Networks}, 
  year={2026},
  volume={28},
  number={},
  pages={3002-3054}
}

\end{document}